# Simultaneous achievement of record-breaking colossal magnetoresistance and angular magnetoresistance in an antiferromagnetic semiconductor EuSe$_2$

Qingxin Dong[1,2=], Pengtao Yang[1,2=], Zhihao Liu[1,2=], Yuzhi Wang[1,2], Ziyi Liu[1,2], Tong Shi[1,3], Zhaoming Tian[3], Jianping Sun[1,2], Yoshiya Uwatoko[4], Quansheng Wu[1,2]*, Genfu Chen[1,2,5]*, Bosen Wang[1,2]* and Jinguang Cheng[1,2]*

[1] *Beijing National Laboratory for Condensed Matter Physics and Institute of Physics, Chinese Academy of Sciences, Beijing 100190, China*
[2] *School of Physical Sciences, University of Chinese Academy of Sciences, Beijing 100049, China*
[3] *Wuhan National High Magnetic Field Center and School of Physics, Huazhong University of Science and Technology, Wuhan 430074, China*
[4] *Institute for Solid State Physics, University of Tokyo, Kashiwanoha 5-1-5, Kashiwa, Chiba 277-8581, Japan*
[5] *Songshan Lake Materials Laboratory, Dongguan, Guangdong 523808, China*

= These authors contributed equally to this work.

*Corresponding authors: bswang@iphy.ac.cn (BSW); quansheng.wu@iphy.ac.cn (QSW); gfchen@iphy.ac.cn (GFC); jgcheng@iphy.ac.cn (JGC)

**Magnetoresistance effect lays the foundation for spintronics, magnetic sensors and hard drives. The pursuit of magnetic materials with colossal magnetoresistance (CMR) and/or angular magnetoresistance (AMR) has attracted enduring research interest and extensive investigations over past decades[1,2]. Here we report on the discovery of field-induced record-breaking CMR of ~ -10$^{14}$ % and AMR ~ 10$^{14}$% achieved simultaneously in an antiferromagnetic rare-earth dichalcogenide EuSe$_2$. Such intriguing observations are attributed to strong magnetic anisotropy and magnetic-field induced antiferromagnetic to ferromagnetic transition of the localized Eu$^{2+}$ spins, which in turn closes the bandgap by lifting the degeneracy of Se-5$p$ bands near Fermi level. Our DFT calculations perfectly replicate the experimental findings based on the Brillouin function and carries transport model. The present work provides a potential simple antiferromagnetic material for achieving angle-sensitive spintronic devices.**

EuSe$_2$ crystallizes in a tetragonal structure (space group: $I4/mcm$) with lattice parameters $a$ = 6.3959(11) Å and $c$ = 7.859(2) Å, respectively. The crystal structure can be viewed as alternatively stacked layers of Eu$^{2+}$ and [Se$_2$]$^{2-}$ dimers along the $c$ axis, Fig. 1(a). Previous studies on EuSe$_2$ proposed a $C$-type antiferromagnetic (AFM) structure below $T_N \approx 8.0$ K, characterized by antiferromagnetically aligned intralayer and ferromagnetically coupled interlayer spins of Eu$^{2+}$[3,4]. This is in sharp contrast to the isostructural EuTe$_2$ adopting an $A$-type AFM order below $T_N \approx 11$ K[5,6]. Recent studies on the EuTe$_2$ single crystals found negative colossal magnetoresistance (CMR) $\sim$ -10$^7$ % and angular magnetoresistance (AMR) $\sim 4 \times 10^4$ %, concomitant with magnetic-field-induced spin re-orientation to a polarized ferromagnetic (FM) state[5,6]. As an isostructural compound, the magneto-transport properties of EuSe$_2$ remain largely unexplored so far. In this work, we have grown high-quality EuSe$_2$ single crystals using self-fluxing method and discovered record-breaking CMR $\sim$ -10$^{14}$ % and AMR $\sim$ 10$^{14}$ % simultaneous achievable at low magnetic field, which is closely related to strong magnetic anisotropy and magnetic-field induced metamagnetic transition of localized Eu$^{2+}$ spins.

**Anisotropic magnetic properties and metamagnetic transition**

As shown in Fig. 1(b), as-grown EuSe$_2$ single crystals are dark black in color, rectangular shape with typical dimensions of $\sim 3 \times 0.5 \times 0.5$ mm$^3$ and its longest dimension along the $c$-axis. These crystals are nearly two orders larger than those in previous report ($\sim 0.18 \times 0.22 \times 0.10$ mm$^3$)[3,4]. The Eu:Se ratio is confirmed to be 1:2 via the energy-dispersive X-ray spectroscopy (EDS) and the element mapping reveals a uniform distribution of Eu and Se in crystals, Extended Data Fig. 1(a, b). Details about the crystallographic data and atomic parameters of EuSe$_2$ via structural refinements are summarized in Extended Data Tables 1-2.

The magnetic properties of EuSe$_2$ were characterized by dc magnetic susceptibility $\chi(T)$ and magnetization $M(B)$ with the magnetic field applied parallel ($B_{//}$) and perpendicular ($B_\perp$) to $c$-axis. As seen in Fig. 1(c), the zero-field-cooled (ZFC) and field-cooled (FC) $\chi_{//}(T)$ curves under $B_{//}$ = 0.1 T are overlapped with each other and exhibit a pronounced cusp-like anomaly at $T_N \approx 8$ K, which shifts to lower temperatures with increasing $B_{//}$ and vanishes at $B_{//} \geq 2$ T. In contrast, the ZFC/FC $\chi_\perp(T)$ curves under $B_\perp$ = 0.1 T display a step-like increment below $T_N$, and change to a kink-like or plateau feature at higher $B_\perp$. These characteristics are consistent with a spontaneous collinear AFM alignment of Eu$^{2+}$ spins oriented along $c$-axis at low $B$, which can be either polarized at $B_{//} \geq 2$ T or reoriented to align with external $B$ at higher $B_\perp$.

Such a scenario is confirmed by the isothermal $M(B)$ curves with $B // c$ and $B \perp c$. As shown in Fig. 1(d), the $M_{//}(B)$ curve at $T$ =2 K < $T_N$ first increase gradually and then experience a rapid enhancement near $B_{//} \approx 1.7$ T before reaching the saturation value of ~7 $\mu_B$/Eu$^{2+}$. The critical magnetic field of $B_{//c} = 2.1$ T at 2 K for the metamagnetic transition can be determined from the interception point shown in Fig. 1(d). In contrast, the development process to the fully polarized FM state for $M_\perp(B)$ is much retarded. As seen in Fig. 1(d), the $M_\perp(B)$ at 2 K experiences an initial quick rise and then undergoes a progressive linear increase until it levels off above a critical $B_{\perp c} \approx$ 4.4 T, which is larger than $B_{//c} = 2.1$ T. As the temperature is raised gradually to above $T_N$, both $M_{//}(B)$ and $M_\perp(B)$ curves evolve smoothly with slight concave curvature at high fields. These above characterizations demonstrate that EuSe$_2$ is a novel binary antiferromagnet with pronounced magnetic anisotropy having an easy axis of localized Eu$^{2+}$ spins along the *c*-axis direction and field-induced metamagnetic transition at moderate fields[3,4].

In the paramagnetic region, the inverse susceptibilities $\chi^{-1}(T)$ above 50 K follows a linear temperature dependence, and a fitting to the modified Curie-Weiss law yields an effective moment $\mu_{eff} = 2.84C^{1/2} \approx 7.9\mu_B$, in good agreement with the theoretical expectation for the localized moment of Eu$^{2+}$, Extended Data Fig. 2(d). The magnetic entropy $S_m(T)$ of EuSe$_2$ associated with the AFM order of localized Eu$^{2+}$ spins is estimated by integrating the magnetic contributions of $C_{mag}/T$, which is obtained by subtracting the electronic and phonon contributions from the total $C(T)$, Extended Data Fig. 3. The obtained $S_m(T)$ increases rapidly and reaches 11.9 J mol$^{-1}$ K$^{-1}$ at $T_N$, then grows slowly to 16.8 J mol$^{-1}$ K$^{-1}$ at 45 K, which is very close to the theoretical value of $R\ln(2S+1) = 17.3$ J mol$^{-1}$ K$^{-1}$ for Eu$^{2+}$ with $S = 7/2$.

## Extraordinarily large CMR effect

The magnetic anisotropy usually endows interesting anisotropic magneto-transport properties that deserve in-depth investigations. To this end, we measured temperature-dependent resistivity $\rho(T)$ ($I//c$) of EuSe$_2$ single crystals under various magnetic fields applied parallel and perpendicular to the *c*-axis. The $\rho(T)$ at zero field exhibits a typical thermally activated semiconducting behavior with an energy gap of $E_g \approx 32$ meV as estimated from a linear fitting to the log$\rho$ vs.1000/$T$, and the analysis of Hall effect reveals dominant hole-type carriers with the concentration of ~10$^{19}$ cm$^{-3}$, Extended Data Fig. 4. Interestingly, the $\rho(T)$ at low temperatures can be significantly reduced by applying magnetic fields along both directions, exhibiting an extraordinarily large CMR effect in EuSe$_2$. Yet, some quantitative differences of $\rho(T)$ are noticeable for $B // c$ and $B \perp c$, Fig.1(e, f).

Unlike $\rho(T)$ at zero field, the $\rho(T)$ at $B_{//} = 1$ T can be measured down to 4 K so that an inflection point is clearly visible at $T_N \approx 8$ K, below which $\rho(T)$ continues to increase. With increasing $B_{//}$ to 1.4 T, the low-temperature upturn in $\rho(T)$ is suppressed markedly and $T_N$ shifts down to ~ 5.5 K. Meanwhile, a broad hump centered around $T^* \approx 13$ K appears at higher temperatures, producing a metallic-like region in $\rho(T)$ between $T_N$ and $T^*$. Such a metallic region is enlarged quickly accompanying the concomitant enhancement of $T^*$ and suppression of $T_N$ upon further increasing $B_{//}$ that continuously suppresses the low-temperature upturn resistivity. Finally, the metallic region is stabilized down to the lowest temperature at $B_{//} \geq 1.7$ T and the magnitude of $\rho(T)$ continuously decreases, resulting in the CMR values as large as $-10^{14}$% at low temperatures under $B_{//} \geq 2$ T. The resistivity hump around $T^*$ can reach ~30 K at 7 T. As mentioned above, the overall behaviors of $\rho(T)$ for $B \perp c$ are similar to those for $B // c$, except that the complete suppression of low-temperature upturn in $\rho(T)$ requires a much higher $B_\perp \geq 4$ T, in line with the anisotropic magnetic properties shown above.

Similar behaviors are reproduced for $\rho(T)$ ($I \perp c$) under $B // c$ and $B \perp c$ configurations for another two samples, Extended Data Fig. 5, confirming an intrinsic behavior of $EuSe_2$. This also implies that the observed CMR effect in $EuSe_2$ is mainly determined by the orientation between $B$ and the $Eu^{2+}$ spins, independent of the current direction. Thus, the intriguing magneto-transport properties of $EuSe_2$ are dictated by the magnetic behaviors of localized $Eu^{2+}$ spins, as found in $EuTe_2$[5,6] and $HgCr_2Se_4$[7,8]. To further elucidate this point, we made a plot of $\rho(B)$ at different temperatures for $B \perp c$ by extracting the data from corresponding $\rho(T)$ curves in Fig. 1(f). As can be seen in Fig. 1(g), the $\rho(B)$ at 2 K decreases quickly and levels off above a critical field of 4.5 T, matching perfectly to the $B_{\perp c}$ determined from $M_\perp(B)$, Fig. 1(d). On the other hand, the $\rho(B)$ curves in the intermediate temperature range $T_N < T < T^*$ exhibit smooth evolutions as $M_\perp(B)$. These results confirm that field-induced metallization in $EuSe_2$ is intimately correlated with the spin-polarized FM state of $Eu^{2+}$ spins above $B_{//c} = 2.1$ T and $B_{\perp c} = 4.4$ T.

Based on the above $\rho(T)$, $C(T)$ and $\chi(T)$ data, we constructed the $T$-$B$ phase diagram of $EuSe_2$, Fig.1(h, i). As $B_\perp$ increases, the AFM order at $T_N$ is suppressed gradually and disappears at ~ 4.5 T; before that, field-induced FM interactions manifested as a broad crossover to metallic behavior in resistivity below $T^*$, which starts to appear above ~ 1.85 T and enhances up ~ 30 K at 7.0 T in a linear manner. Accordingly, pronounced negative CMR appears accompanied with the field-induced transition from an AFM semiconductor to a FM metal. Similar phase diagram is obtained for $B // c$. The color contour maps of $\log(\rho)$ superimposed on the phase diagram further highlight the intimate correlations between the magnetic and magneto-transport behaviors in $EuSe_2$.

## Significant AMR effect

In addition to the extraordinary CMR effect, the distinct values of $B_{//c} = 2.1$ T versus $B_{\perp c} = 4.4$ T also produce significant AMR in $EuSe_2$. Figure 2(a, b) show the $\rho(T)$ ($I//c$) at fixed $B = 2.5$ T and 5.5 T by rotating the magnetic field from $\theta = 0°$ ($B \perp c$) to $\theta = 90°$ ($B // c$). For $B_{//c} < B = 2.5$ T $< B_{\perp c}$, the $\rho(T)$ at low temperatures is reduced remarkedly as the field rotates from $\theta = 0°$ to 90°, in reminiscent of the $\rho(T)$ by varying $B_\perp$ in Fig. 1(f), leading to an extremely large AMR defined as [$\rho(\theta = 0°)$-$\rho(\theta = 90°)$]/$\rho(\theta = 90°)$*100% below $T_N$. As the temperature increases, the AMR quickly reduces in the temperature range $T_N < T < T^*$ and approaches zero at $T > T^*$. In contrast, for $B = 5.5$ T larger than both $B_{//c}$ and $B_{\perp c}$, the sample retains in the field-induced spin-polarized FM and metallic state below $T^*$ and the AMR is significantly reduced, Fig. 2(b).

Such an AMR effect can be visualized by plotting angle-dependent resistivity $\rho(\theta)$ at various fixed $B$ and 2 K, Fig. 2(c). As the field increases from 1.25 T to 1.75 T, although $\rho(\theta)$ data are only partially available owing to the measurement limit, the magnitude of AMR spans 3-10 orders and increases with field, far exceeding typical AMR materials as discussed below[2,9,10]. For a fixed $B = 2$ T, a dramatic change of resistivity yields an AMR $\sim 10^{14}$% from $\theta = 0°$ to $\theta = 90°$. As the field increases further, the AMR is gradually suppressed but remains larger than other typical materials [9-12]. These results demonstrate that the magneto-transport properties of EuSe$_2$ are mainly dictated by the relative orientation of $B$ with respect to the easy $c$-axis of Eu$^{2+}$ spins. Such a two-fold symmetry can be vividly illustrated in the polar plots of log$\rho(\theta)$ data, Fig. 2(e) for $B < 4.0$ T, above which the fully polarized FM state gives an isotropic log$\rho(\theta)$.

Figure 2(d) shows the $\rho(\theta)$ under a fixed $B = 2$ T at various temperatures up to 20 K. The anisotropy decreases sharply once above $T_N \approx 8$ K. Accordingly, the AMR changes from two-fold symmetry at $T < T_N$ to nearly isotropic at 10-20 K, Fig. 2(f). The AMR decreases rapidly with increasing temperature, but it retains a moderate value of $\sim 130$% at 20 K, higher than typical AMR materials[1,2,6,10,13-16] due to enhanced magnetic fluctuations.

## Comparisons with other CMR/AMR materials

The above characterizations have demonstrated the simultaneous achievement of CMR and AMR in the EuSe$_2$ single crystal with unprecedent magnitude of $\sim 10^{14}$% at low $B = 2$-3 T. Figure 3 compare the CMR/AMR values of EuSe$_2$ with typical CMR and AMR materials[1,6,7,14,17-30]. As seen in Fig. 3(a), EuSe$_2$ is prominently located on the top of diagram with the maximum CMR far beyond other typical CMR materials, including the perovskite manganites, the Eu-based magnetic materials such as EuMn(Bi,Sb)$_2$, EuCd$_2$P$_2$, and EuTe$_2$, as well as the FM semiconductor Mn$_3$Si$_2$Te$_6$ with the topological nodal-line degeneracy of spin-polarized bands, thus setting a new record of CMR. In addition to an extraordinary AMR via comparison with other typical AMR materials[1,2,6,10,13-16], Fig. 3(b), EuSe$_2$ also displays strong tunability of AMR by magnetic field from ~$10^{14}$ % to ~$10^2$ %, as recently found in Mn$_3$Si$_2$Te$_6$ with an extremely large AMR up to ~$10^{11}$ %[1]. Here we further raised the experimental record of AMR to ~$10^{14}$ % in EuSe$_2$. The present work thus demonstrates a concurrent achievement of the highest CMR and AMR values in EuSe$_2$, which may offer an excellent material platform to realize extremely sensitive angle-dependent functionalities.

## Theoretical calculations

To further understand the magneto-transport properties of EuSe$_2$, especially the extraordinary CMR and AMR, its electronic structures were calculated, Fig. 4(a,b) and Extended Data Fig. 6 (a-d). For a *C*-type AFM state along the $\hat{z}$ direction, the indirect bandgap (~ 0.16 eV) between the Se-p orbits emerges along the M-Γ line, Fig. 4(a), which is consistent with the narrow-gap semiconducting behavior at zero field. For the FM configuration along the $\hat{x}$ (or $\hat{z}$) direction, the Fermi surface contains four small electron pockets located around the Γ point with a $C_4$ symmetry and one hole pocket centered at the Γ point. The applied magnetic field induces spin splitting of the Se-5*p* energy bands, thereby pushing up valence band and pulling down conduction band, resulting in a closure of bandgap and metallic state. In this regard, the observed CMR or field-induced semiconducting-metallic transition in EuSe$_2$ is attributed to the regulation of Fermi surfaces. Meanwhile, the same electronic structures for the FM configuration along the $\hat{x}$ (or $\hat{z}$) direction are also consistent with the fact that the observed CMR does not depend on the direction of the applied field with respect to the *c* axis. With the magnetic field up to ~ 2.0-4.5 T, the magnetization reaches saturation only for *B*//*c* and the band gap closes; while the band gap still retains for *B*⊥*c*. Owing to such difference in the critical fields, EuSe$_2$ undergoes a transition from an insulating state to a highly conductive state as the field rotates from *B*⊥*c* to *B*//*c*, reproducing the observed AMR. To quantitatively reproduce the experimental results, the magnetization curves were constructed based on the Brillouin function of paramagnets, Fig. 4(c) and Extended Data Fig. 7, but replace the *B* with $f(B)$:

$$\frac{M}{M_s} = \frac{2J+1}{2J} \coth\left(\frac{2J+1}{2J}\alpha\right) - \frac{1}{2J}\coth\frac{\alpha}{2J}, \quad M_s = g_J J \mu_B \quad (1)$$

$$\alpha = \frac{J g_J \mu_B f(B)}{k_B T} \quad (2)$$

$$f(B) = A\left(\frac{1}{1+e^{-\frac{B}{\alpha T}}} - 0.5\right) + sign(B)\frac{C|B|^\beta}{D+|B|^\beta} + B \quad (3)$$

where the *J*, $g_J$, $\mu_B$, *T*, *B* are the total angular momentum, Lande *g* factor, Bohr magneton, Boltzmann constant, temperature and magnetic field, respectively. Meanwhile, we use the following expressions to describe this magnetization dependency of $E_{CBM}$ and $E_{VBM}$, which are magnetization dependent

$$E_{CBM} = \frac{E_{gap}}{2} - G\left(\frac{M(B,T)}{M_S}\right), E_{VBM} = -\frac{E_{gap}}{2} + G\left(\frac{M(B,T)}{M_S}\right) \quad (4)$$

The $E_{gap}$ is the gap of a AFM state, $M(B, T)$ is the field- and temperature-dependent magnetization with the $M_S$ being its saturated value, and $G$ is a monotone increasing function. A linear magnetization-dependent change in both the $E_{CBM}$ and $E_{VBM}$ is adopted, with $G(M(B, T)/M_S) = 0.5 \times (E_{gap} + E_{overlap}) \times (M(B, T)/M_S)$, where the $E_{overlap}$ is the energy overlap of the upper and lower band. Such a setup not only satisfies the size of the initial energy gap, but causes an insulator-metal phase transition for $B > B_c$, which closes the energy bandgap in Fig. 4(b). And then the simulated resistivity $\rho(T, B)$ obtained based on the above expressions are in Fig. 4(d). All the $M(T, B)$ and $\rho(T, B)$ behaviors perfectly fit both theoretically and experimentally, validating the validity of the above proposed mechanism.

## Discussions

The pursuit of CMR/AMR materials has been a subject of intense research in spintronic physics and materials science owing to the intrinsic mechanisms and potential applications in magnetic memory devices and sensors [31,32]. Several CMR/AMR candidates with different mechanisms have been discovered. One example is the ferrimagnetic insulating materials and layered magnetic artificial structures[33-35], as exemplified by the perovskite-type manganese oxides[10,34], where the CMR over several orders of magnitude was achieved near the double-exchange-interaction driven FM and insulating-metal transition. The other candidates are magnetic topological materials, whose properties are strongly influenced by topological electronic structures coupled with adjustable spin configurations[1,16,33]. In a ferrimagnetic semiconductor $Mn_3Si_2Te_6$, the CMR/AMR originates from non-trivial topological band degeneracy driven by chiral molecular orbital states, which can be lifted by the spin orientation leading to a metal-insulator transition in a FM phase. However, many of these CMR/AMR ferrimagnets generally require high fields and thus have disadvantages of high energy consumption and poor stability, which limits the large-scale applications.

On the contrary, collinear antiferromagnets have zero net magnetic moment. It implies that information stored in the AFM moments would be invisible to magnetic probes, insensitive to disturbing fields, and the AFM element wouldn't magnetically affect its neighbors, regardless of how densely the elements are arranged. In addition, the intrinsic high frequencies of AFM dynamics and large magneto-transport effects make antiferromagnets distinct from ferromagnets. In this scenario, an alternative proposal has been put forward to construct fast-switching and high-density spintronic devices in antiferromagnets[9,12,34,35]. Therefore, achieving CMR/AMR in antiferromagnets is greatly valuable and has promising application prospects[36-38]. In this work, the discovery of a record-breaking CMR ~ -$10^{14}$% and AMR ~ $10^{14}$% at lower magnetic fields in EuSe$_2$ provides a novel material platform for in-depth explorations of CMR/AMR antiferromagnets for angle-sensitive spintronic devices.

## Conclusions

In summary, we discovered the simultaneous achievement of record-breaking values of CMR ~ -$10^{14}$% and AMR ~ $10^{14}$% at lower magnetic fields in a novel binary AFM semiconductor EuSe$_2$. These observations are attributed to the magnetic anisotropy and field-induced metamagnetic transition of magnetic ground state that modifies the electronic band structures near Fermi level by closing the bandgap. The experimentally observed $\rho(T, B)$ and CMR were perfectly replicated by constructing the Brillouin function and carrier transport model according to the DFT calculations. Our present work provides a novel material platform for angle-sensitive spintronic devices, and offers fundamental insights into magneto-transport of Eu-based antiferromagnets.

## Methods

### Single-crystal growth and characterization

Single crystals of EuSe$_2$ were grown by self-fluxing method[3,4]. The raw materials Eu, Se, and Li$_2$Se were mixed in a molar ratio of ~1:8:1, placed into a graphite crucible and then sealed in an evacuated quartz tube. The tube was slowly heated to 540 °C within 24 h in a box furnace, held at 540 °C for 96 h and then cooled to 50 °C at a rate of 2.0 °C/h. The brick-shaped single crystals with typical size of ~ 3.0×0.40×0.30 mm$^3$ can be obtained by washing the reacted product with ether and ethanol sequentially. Single-crystal X-ray diffraction (SXRD) were conducted on a BRUKER D8 VENTURE single-crystal diffractometer at 273 K. The XRD pattern was well refined to extract the atomic coordinates and thermal displacement parameters of EuSe$_2$. The energy-dispersive X-ray spectroscopy (EDS) equipped with a Phenom scanning electron microscope was employed to determine the chemical compositions.

### Physical property measurements

Electrical resistivity $\rho(T)$ and Hall resistivity $\rho_{xy}(B)$ were measured by the standard four-probe method and five-probe method, respectively, using the Physical Property Measurement System (PPMS-9T, Quantum Design) (0 ≤ H ≤ 9.0 T, 1.8 ≤ T ≤ 400 K). For $\rho(T)$ measurements, the current is applied along the c-axis (I//c) with the magnetic fields parallel ($B//c$) and perpendicular ($B \perp c$) to the $c$-axis, respectively. The specific heat $C(T)$ under magnetic fields up to 4.0 T was measured in PPSM by two-tau relaxation method. The magnetization at fixed magnetic fields and temperatures $M(T, B)$ was measured with a Magnetic Property Measurement System (MPMS-7T, Quantum Design) ($0 \leq H \leq 7.0$ T, $1.8 \leq T \leq 300$ K).

**First-principles calculations**

The band structures of EuSe$_2$ was calculated by using Vienna *ab* initio simulation package (VASP)[39] with the PBESOL exchange-correlation potential considering the spin-orbit coupling[40]. The relaxed crystal structure with lattice constants of $a = b = 6.294$ Å and $c = 7.779$ Å was adopted in the DFT calculations. The localized Wannier functions for the $4f$ and $5d$ orbitals of Eu$^{2+}$, the Se-$4p$ orbitals are used as the basis to construct the tight-binding Hamiltonian using the Wannier 90 package[41]. The anomalous Hall conductivity is calculated using the Wannier Tools package[42].

**Acknowledgements**

We thank Yuxin Wang, Kun Jiang and Jiangping Hu for fruitful discussions. This work is supported by the National Key R&D Program of China (2023YFA1607400, 2023YFA1406100), the National Natural Science Foundation of China (12025408, 12174424, 12274436, 11874400, 11921004, 11888101, 12188101,12274440), the Strategic Priority Research Program of CAS (XDB33000000), Chinese Academy of Sciences President's International Fellowship Initiative (2024PG0003), and the Outstanding member of Youth Promotion Association of CAS (Y2022004). This work was carried out at the Synergetic Extreme Condition User Facility (SECUF).

**Competing interests**

The authors declare no competing interests.

**Author contributions**

J.G.C. and B.S.W. supervised this project; Q.X.D and P.T.Y synthesized the EuSe$_2$ single crystals and characterized their structure via XRD and EDS; Q.X.D., P.T.Y, Z.Y.L. performed electrical resistivity, Hall coefficient, magnetic susceptibility, specific heat measurements, and data analyses together with G.F.C, B.S.W. and J.G.C.; Y.U., J.P.S. T.S., Z.M.T. an B.S.W. gave some useful comments and advises for measurements; Z.H.L, Y. Z. W. and Q.S.W. conducted the first principles calculations and gave good advice from a theoretical perspective. J.G.C., B.S.W., G.F.C., Q.S.W., Q.X.D., P.T.Y. and Z.H.L wrote the paper with inputs from all coauthors.

**Data Availability**

All data are available from the corresponding authors upon reasonable request. Source data are provided with this paper.

## Online content:
Any methods, additional references, Nature Portfolio reporting summaries, source data, extended data, supplementary information, acknowledgements, peer review information; details of author contributions and competing interests; and statements of data and code availability are available at XXX.

**Figures and captions**

**Figure. 1 Physical properties and CMR effect of EuSe$_2$ single crystals.** (a) Crystal structures of EuSe$_2$ with an AFM configuration. (b) X-ray pattern of EuSe$_2$ with (110) plane. Inset shows the as-grown single crystal. (c) $\chi_{//}(T)$ and $\chi_{\perp}(T)$ below 30 K at 0.1 T under field-cooling (FC) and zero-field-cooling (ZFC) processes, respectively. (d) The isothermal $M(B)$s for both $B//c$ and $B \perp c$, respectively, at 2 K and 30 K. (e, f) $\rho(T)$ of EuSe$_2$ with the $B//c$ and $B \perp c$, respectively. (g) $\rho(H)$ at various temperatures for $B \perp c$. (h, i) Temperature-field diagram of EuSe$_2$ for $B//c$ and $B \perp c$, respectively based on $\rho(T)$, $\chi(T)$, and $C(T)$ data. The color contour plot indicates the absolute log($\rho$) values.

**Figure. 2 AMR effect of EuSe$_2$.** (a, b) The $\rho(T)$ of EuSe$_2$ with various angles under two fixed magnetic fields of $B$ = 2.5 T and 5.5 T, respectively. (c, d) Angular-dependent resistivity, $\rho(\theta)$, for EuSe$_2$ under various fields at 2 K and $\rho(\theta)$ at various temperatures under a fixed field of 2.0 T. (e, f) Polar plots of the log-scaled resistivity at a fixed temperature of 2 K and a fixed field of 2.0 T, respectively.

**Figure. 3 Comparison of the CMR and AMR with other typical materials.** (a) The CMR of EuSe$_2$ and other well-known materials, including perovskite manganites, Eu-based compounds and topological electronic materials. (b) The AMR of EuSe$_2$ under various fields at 2.0 K compared with other representative magnetic materials.

**Figure. 4 DFT calculations and numerical simulations of EuSe$_2$.** (a, b) The enlarged band structures of the $C$-type AFM and FM configurations, respectively, showing an energy gap of ~ 0.16 eV in an AFM state and a closure of bandgap in a FM state. (c, d) The numerical simulations of magnetization $M(T, B)$ for $B \perp c$ based on the Brillouin function and the electrical resistivity $\rho(T, B)$ under various fields based on the carrier transport model, both are consistent with the experimental observations.

# Figures

## Figure 1

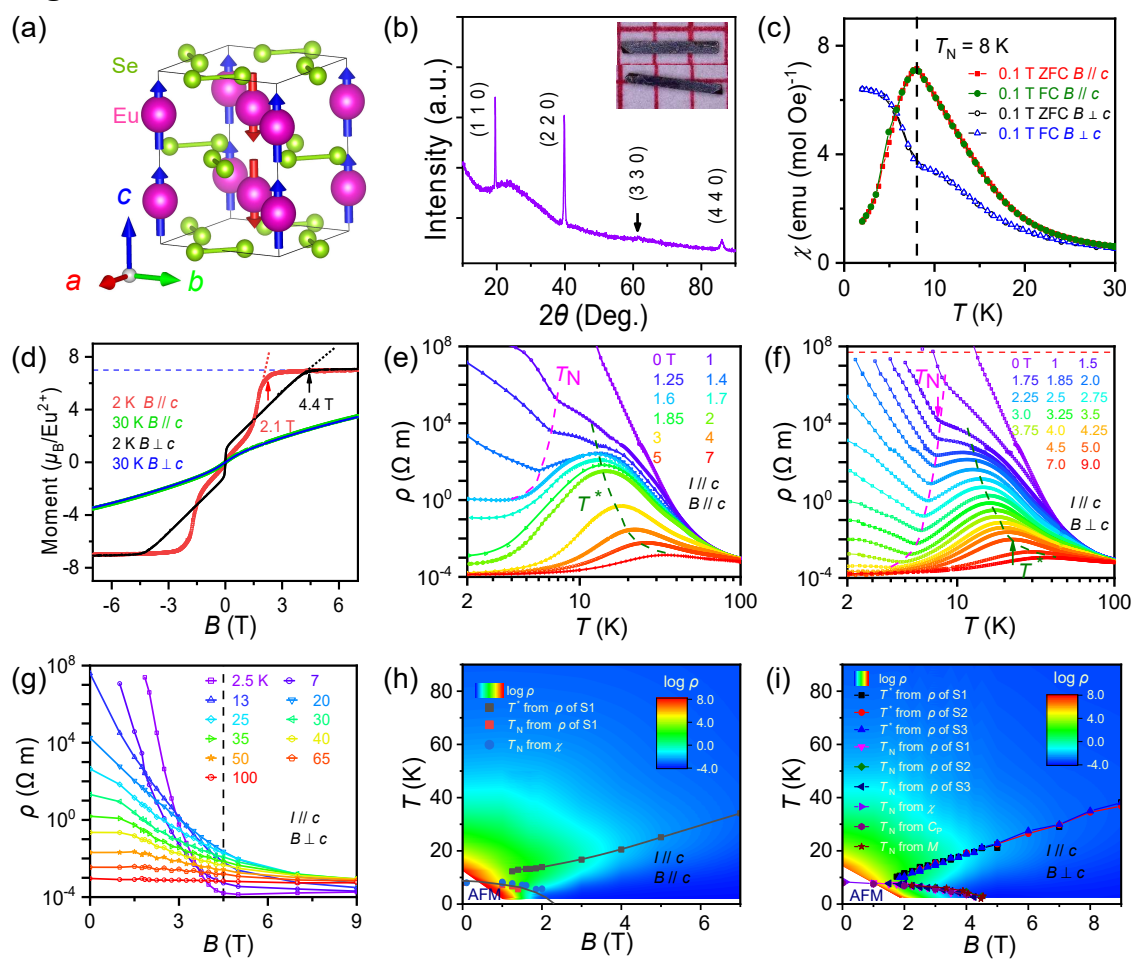

**Figure 2**

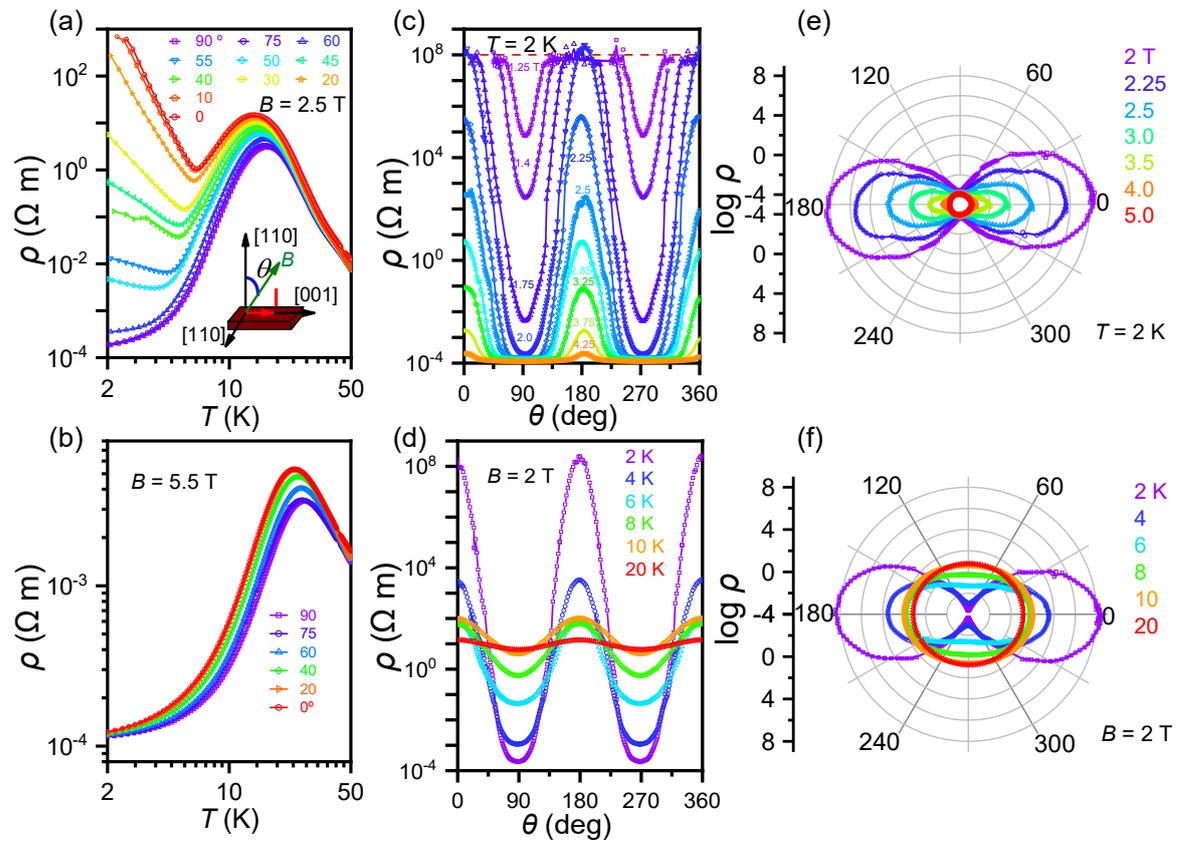

**Figure 3**

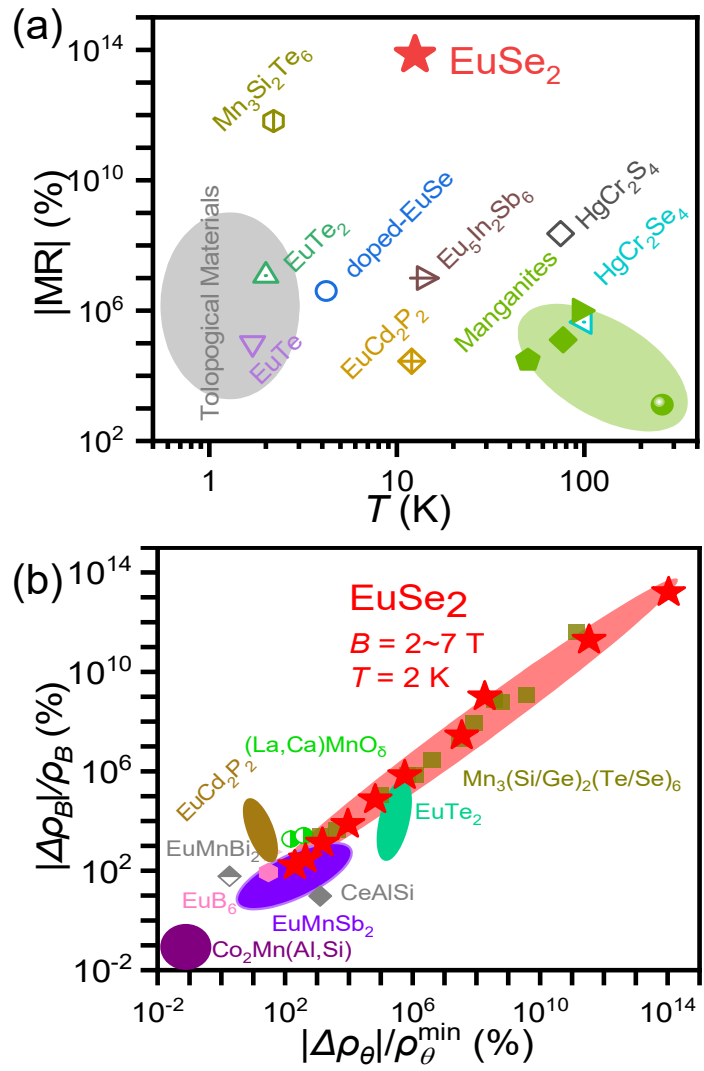

**Figure 4**

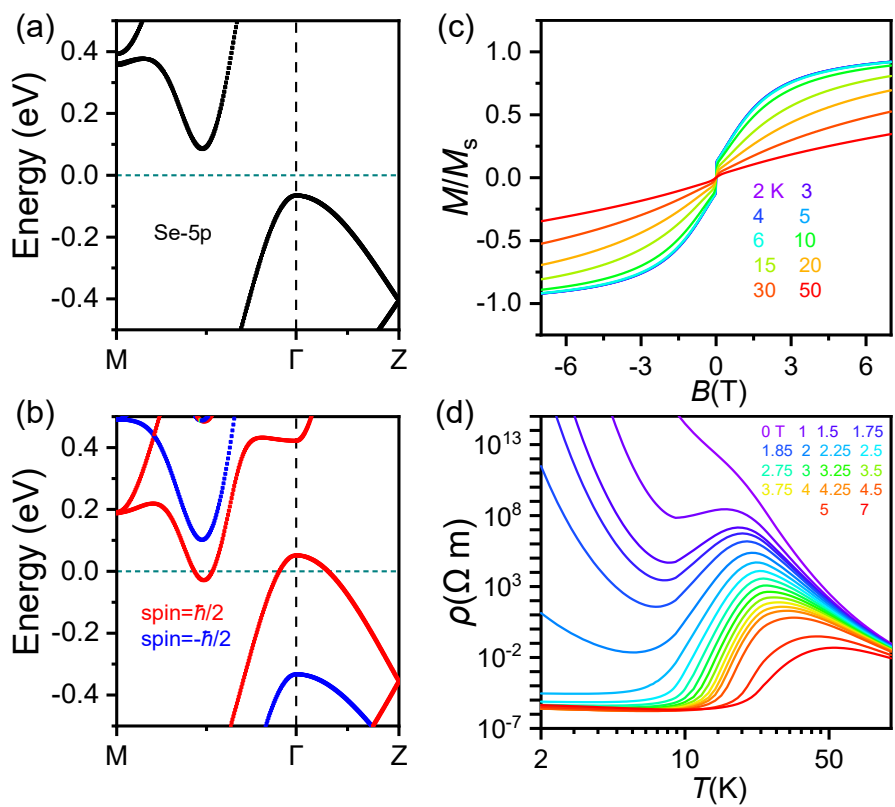